\documentclass{pasa}%

\title[Photometric asymmetry]{Photometric asymmetry between clockwise and counterclockwise spiral galaxies in SDSS}
\author[Shamir]{Lior Shamir$^1$  % \thanks{This is an example of author footnote}\\
\affil{$^1$Lawrence Technological University, 21000 W Ten Mile Rd, Southfield, MI 48075, USA}%
}
\jid{PASA}
\doi{10.1017/pas.\the\year.xxx}
\jyear{\the\year}

\usepackage[authoryear]{natbib}
\bibpunct{(}{)}{;}{a}{}{,}
\usepackage{aas_macros}
\usepackage{hyperref} 
\usepackage{epstopdf}
\hypersetup{colorlinks,citecolor=blue,linkcolor=blue,urlcolor=blue}

\begin{document}%
\begin{abstract}
While galaxies with clockwise and counterclockwise handedness are visually different, they are expected to be symmetric in all of their other characteristics. Previous experiments using both manual analysis and machine vision have shown that the handedness of Sloan Digital Sky Survey (SDSS) galaxies can be predicted with accuracy significantly higher than mere chance using its photometric data alone, showing that the photometric pipeline of SDSS is sensitive to the handedness of the galaxy. However, some of these previous experiments were based on manually classified galaxies, and the results may therefore be subjected to bias originated from the human perception. This paper describes an experiment based on a set of 162,514 celestial objects classified as clockwise and counterclockwise spiral galaxies in a fully automatic process, showing that the source of the asymmetry in SDSS database is not the human perception bias. The results are compared to two smaller datasets, and confirm the observation that the handedness of galaxies in SDSS database can be predicted by the way SDSS measures their photometry, and show that the position angle of counterclockwise galaxies computed by SDSS photometry pipeline is consistently higher than the position angle computed for galaxies with clockwise patterns. The experiment also shows statistically significant differences in the measured magnitude of SDSS galaxies, according which galaxies with clockwise patterns imaged by SDSS are brighter than galaxies with counterclockwise patterns. The magnitude of that difference changes across RA ranges, and exhibits a strong correlation with the cosine of the right ascension.
\end{abstract}
\maketitle%

\section{INTRODUCTION}
\label{introduction}

%While it is clear that spiral galaxies rotate, the physical nature of the rotation is not yet fully understood. Observations have shown that the physics of galaxy rotation is not driven solely by the gravity of the visible matter \citep{rubin1970rotation}, and possible explanations include the effect of dark matter halos \citep{navarro1996structure} or the modifications in Newton's laws of universal gravitation \citep{milgrom1983modification}.

While galaxies with clockwise handedness are expected to be symmetric to galaxies with counterclockwise handedness, previous experiments \citep{sha16} have shown that clockwise galaxies imaged by Sloan Digital Sky Survey \citep{york2000sloan} are photometrically different from galaxies with counterclockwise patterns. The experiment was done by separating a population of spiral galaxies into clockwise and counterclockwise galaxies, and then collecting the photometric information of each galaxy from SDSS Catalog Archive Server (CAS). That was done by using the Galaxy Zoo 2 \citep{willett2013galaxy} galaxies classified manually as spirals, as well as with another dataset of 10,281 galaxies classified as clockwise and counterclockwise spiral galaxies in a fully automatic process, and without human intervention.

Then, supervised machine learning was trained using these photometric data such that the label of each class was the handedness. Experimental results show that by using the photometric information the handedness of the galaxy (clockwise or counterclockwise) can be predicted in accuracy of $\sim$64\%, which is much higher than random guessing accuracy of 50\% (P$<10^{-5}$). The experiment using the dataset of galaxies that were annotated in a fully automatic process provided a comparable handedness prediction accuracy of $\sim$65\%. The analysis also revealed that several different photometric measurements such as the SDSS `Stokes U' parameter exhibit a statistically significant difference between clockwise and counterclockwise galaxies \citep{sha16}.

Other observations related to handedness asymmetry measured the number of clockwise and counterclockwise galaxies, showing evidence of asymmetry between the number of clockwise and counterclockwise galaxies. \cite{land2008galaxy} used a large dataset of galaxies annotated by crowdsourcing, showing that after correcting for the substantial human bias the number of galaxies with clockwise handedness was higher than the number of counterclockwise galaxies, but the difference was not statistically significant \citep{land2008galaxy}. A more recent analysis using 13,440 automatically classified galaxies \citep{sha16} showed a higher number of galaxies with clockwise handedness compared to galaxies with counterclockwise handedness, which is also in agreement with the higher number of Galaxy Zoo \citep{lintott2011galaxy} clockwise galaxies observed in \citep{shamir2012handedness}. On the other hand, the dataset of 10,281 galaxies used in \citep{sha16} showed no statistically significant preference \citep{sha16}.

Since most of these studies were done by using manual classification of the galaxies, a possible explanation is that the asymmetry is driven by a bias in the human perception. This paper describes an experiment that uses data annotated in a fully automatic process, and with no human intervention that can induce bias. Since the dataset is far larger than previous datasets used for that purpose, it can also be used to show statistically significant differences between measurements of galaxies with clockwise patterns and galaxies with counterclockwise patterns.

\section{DATA}
\label{computer_dataset}

The galaxies used in the experiment were galaxies classified as spiral galaxies in the catalogue of broad morphology of $\sim$3,000,000 SDSS Data Release 8 galaxies \citep{kum16}, which was generated automatically by applying the Wndchrm image classifier \citep{shamir2008source,shamir2013wnd} to the galaxy images \citep{shamir2009automatic,kuminski2014combining}. The initial set of $\sim$3,000,000 galaxies was selected such that all galaxies had a Petrosian radius (measured on the r band) of at least 5.5'', the Petrosian radius error was less than 5'', and the flags were selected such that none of the objects was identified as ``bad sky'', ``bad radial'', ``too large'', too close the the edge of the frame, or had more than one peak or Petrosian radius \citep{kum16}. These constraints provided a set of galaxies with identifiable morphology, allowing their separation into spiral and elliptical galaxies. That was done by applying an automatic classifier that can analyze galaxy images and annotate them with their broad morphological types \citep{shamir2009automatic,kuminski2014combining}. Each galaxy was assigned with its broad morphological type of early or late type, but also with the certainty value within the interval (0,1) that the classification is correct. A certainty value close to 0.5 means that the certainty of the classification is nearly random, while a certainty value close to 1 indicates that there is a very low chance that the annotation is incorrect. A detailed description of the catalogue is available in \citep{kum16}.

Galaxies that were classified as spiral galaxies with certainty higher than 0.54 were used in the experiment, providing a dataset of 740,908 spiral galaxies \citep{kum16}. To assess the consistency of the galaxy catalogue compared to manual annotation of the galaxies, the galaxies were compared to the manual annotation of {\it Galaxy Zoo}. That was done by identifying all galaxies that are included in the catalogue and were also classified by {\it Galaxy Zoo} as debiased ``superclean'', and comparing the morphological annotation of the catalogue to the morphological annotation of the superclean {\it Galaxy Zoo}. Statistical analysis of the 45,377 galaxies included in the catalogue and were also classified by {\it Galaxy Zoo} as debiased ``superclean'' showed that in $\sim$98\% of the cases these galaxies are also identified as spiral by the debiased ``supercleasn'' {\it Galaxy Zoo} dataset \citep{kum16}, showing that the set of spiral galaxies is reasonably accurate for the purpose of identifying spiral galaxies in SDSS. 

Similarly to \citep{sha16}, the spiral galaxies were separated by their handedness to clockwise and counterclockwise galaxies using the {\it Ganalyzer} algorithm \citep{shamir2011ganalyzer,ganalyzer_ascl}, which converts each galaxy image to its radial intensity plot, and then applies automatic peak detection to find and group the peaks along the horizontal line of the radial intensity plot \citep{shamir2011ganalyzer}. Since the galaxy arms are brighter than non-arm pixels at the same radial distance, the peaks are expected to identify the arms. Linear regression is applied across the vertical lines of the radial intensity plot for each group of peaks, and the sign of the slopes reflect the direction of the arm, therefore determining the handedness of the galaxy. The algorithm is described in details and numerous examples in \citep{shamir2011ganalyzer,hoehn2014characteristics,shamir2012handedness}, and the process of galaxy classification is described in \citep{sha16}. 

%The {\it Ganalyzer} algorithm is deterministic, and is not expected to introduce a bias in the form of preference of a certain handedness over the other. To test for the existence of such bias experimentally, {\it Ganalyzer} was applied in the same way described in \citep{sha16,hoehn2014characteristics,shamir2012handedness} to the dataset of manually classified galaxies used in \citep{sha16}. The results showed that 1812 of 5,500 counterclockwise galaxies were correctly classified as counterclockwise, while 1779 of the 5,500 clockwise galaxies were classified correctly as clockwise. The rate of correct classification is $\sim$0.329 among the counterclockwise galaxies, while it is $\sim$0.323 among clockwise galaxies. The t-test of the difference falls within the statistical error $(P\simeq0.503)$. The number of incorrectly classified galaxies in both classes was 24, exhibiting an error rate of 0.0048.

The process of the separation of the galaxies to clockwise and counterclockwise described above provided a dataset of 82,244 galaxies with clockwise handedness and 80,272 galaxies with counterclockwise handedness. The remaining galaxies did not have a clear handedness determined by Ganalyzer, and were therefore excluded from the experiment.

As the results show, the number of galaxies with clockwise patterns is higher than the number of galaxies with counterclockwise patterns. Assuming random 0.5 probability of the galaxy to have each of the two possible patterns, the probability to have such separation by chance can be computed using cumulative binomial distribution, such that the number of tests is 162,516 and the probability of success is 0.5. The probability to have 82,244 or more successes is $P\simeq5\cdot10^{-7}$.

The higher number of clockwise galaxies is aligned with the ratio between clockwise and counterclockwise galaxies in datasets of manually classified spiral galaxies \citep{sha16,shamir2012handedness,hoehn2014characteristics}. Other experiments using automatic \citep{sha16} and manually \citep{land2008galaxy} classified galaxies showed a higher number of galaxies with clockwise handedness, although the difference was not statistically significant.

After the galaxies were classified, 400 random galaxies classified by Ganalyzer as clockwise and 400 random galaxies classified by Ganalyzer as counterclockwise were examined manually to test the consistency of the dataset. Twenty four galaxies classified as clockwise had no clear identifiable handedness, as well as 21 galaxies that were classified by Ganalyzer as counterclockwise. However, none of the galaxies that were examined was clearly misclassified. 

\begin{figure*}[ht]
%\figurenum{<text>}
\includegraphics[scale=0.5]{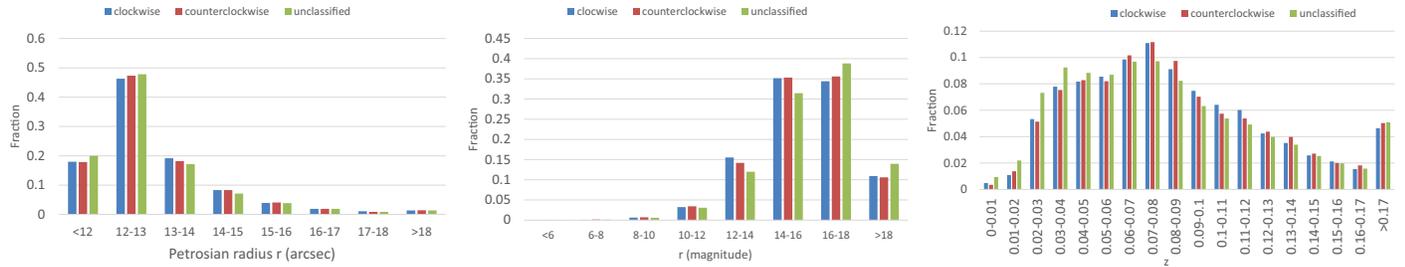}
%\plottwo{<epsfile>}{<epsfile>}
\caption{Distribution of the r magnitude, Petrosian radius measured in the r band, and the distribution of redshift. The distribution of magnitude and radius was measured with the entire dataset, while the distribution of the redshift is among a subset of 10,281 galaxies with spectra.}
\label{distribution}
\end{figure*}

Figure~\ref{distribution} shows the distribution of the r magnitude, Petrosian radius measured in the r band, and the redshift of the galaxies classified by Ganalyzer as clockwise, counterclockwise, and galaxies that could not be classified to any of these classes and remained unclassified. The vast majority of the galaxies do not have spectra, and therefore just 10,281 galaxies that had redshift information could be used. 

As the figure shows, while galaxies with higher r magnitude tend to be classified less frequently into clockwise or counterclockwise galaxies, the distribution of the galaxies that could not be classified by {\it Ganalyzer} is largely aligned with the distribution of the galaxies that were classified as clockwise or counterclockwise.

\section{ANALYSIS AND RESULTS}
\label{results}

Since galaxies are described by multiple measurements rather than a single parameter \citep{brosche1973manifold,djorgovski1987fundamental}, all of the 509 photometric variables from SDSS Data Release 8 \citep{aihara2011eighth} PhotoObjAll table were used for each galaxy in the dataset. As in \citep{sha16}, the galaxies were classified automatically by four different supervised machine learning algorithms: Random Forest \citep{breiman2001random}, Decision Table \citep{kohavi1995power}, Ensembles of Balanced Nested Dichotomies \citep{dong2005ensembles}, and Bagging \citep{breiman1996bagging}. The features of each galaxy were the PhotoObjAll photometric variables, and the class label of each galaxy was its handedness. That is, the purpose of the supervised machine learning was to identify the handedness of a galaxy based on its DR8 photometric variables. Assuming no link between the photometry of the galaxy and its handedness, the prediction accuracy is expected to be equal to random guessing, which would provide 50\% classification accuracy.

Some SDSS galaxies have photometric values such as -9999, which are actually flags and not actual photometric measurements. Since the purpose of the experiment is to analyze the photometry of galaxies, galaxies with photometric values such as -9999 were ignored.

\begin{figure}[ht]
%\figurenum{<text>}
\includegraphics[scale=0.7]{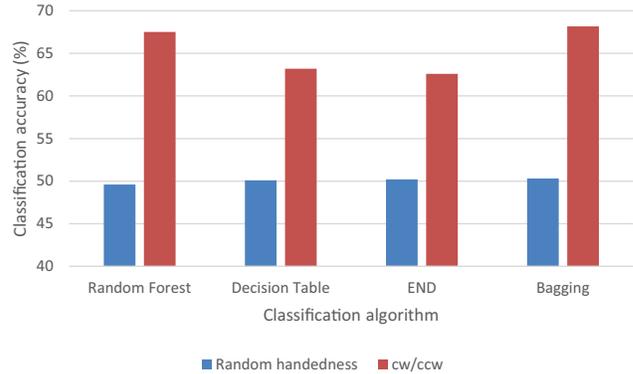}
%\plottwo{<epsfile>}{<epsfile>}
\caption{Prediction accuracy of the handedness using the photometric information. The graph also shows the prediction accuracy when the handedness of each galaxy was replaced with a random handedness.}
\label{computer_accuracy}
\end{figure}

The machine learning was done using the open source Waikato Environment for Knowledge Analysis (WEKA) software \citep{frank2004data,hall2009weka}. The experiments were performed such that 80\% of the galaxies were used for training, and the remaining 20\% were used for testing the ability of the classifier to correctly identify the handedness of the galaxy based on its photometric variables. The classification accuracy was determined by the number of galaxies which their handedness was predicted correctly, divided by the total number of galaxies. Additionally, the same experiments were repeated such that random handedness was assigned to each galaxy. The results are displayed in Figure~\ref{computer_accuracy}.

The graph clearly shows that the photometric variables computed by SDSS contain information related to the handedness, and that the handedness of the galaxy can be predicted by using its SDSS photometric variables. When a random handedness is assigned to each galaxy, the classification accuracy drops to the $\sim$50\% mere chance accuracy. Differences in the classification accuracy achieved by each of the machine learning algorithms are expected, as not all machine learning algorithms are equally powerful, and different algorithms may perform differently on different datasets.

To achieve classification accuracy of 68.15\% as was achieved using the Bagging classification method, 22,151 galaxies should be classified correctly out of the total of 32,503 galaxies used for testing. The probability for 22,151 galaxies to be classified accurately by chance can be computed using cumulative binomial probability such that the number of tests is 32,503, the minimum number of successes is 22,151, and the probability of success in each test is 0.5. The probability for 22,151 or more successes is extremely low ($P<10^{-5}$). These results are in agreement with previous studies using manually classified data or smaller sets of automatically classified galaxies, in which comparable accuracy of $\sim$65\% was observed for the automatic prediction of the handedness of the galaxy \citep{sha16}.

%The prediction of the handedness was done using all variables in SDSS DR8 PhotoObjAll table. To identify the variables that were used for the prediction of the handedness, several feature selection algorithms were used. These algorithms included  Consistency Subset Eval \citep{Liu1996}, Combined Feature Selection (CFS), Filtered Attribute Eval, and Subset Eval \citep{Hall1998}. The variables that were selected by the different methods are shown in Table~\ref{dr8_features}. 

%\begin{table}[ht]
%\begin{center}
%\caption{The variables selected by applying three different automatic feature selection methods: Consistency Subset Eval, Filtered Attribute Eval, and  Combined Feature Selection.}
%\label{dr8_features}
%{ \scriptsize 
%\begin{tabular}{lcc}
%\tableline\tableline
%Consistency    & Combined              & Filtered           \\
%\tableline
% IsoPhiGrad\_u & isoRowcErr\_z & IsoPhiGrad\_g \\
     
% \tableline
%\end{tabular}
%}
%\end{center}
%\end{table}

%\normalsize 

%?????table for the feature selection?????
%?????explain about the variables??????

The prediction of the handedness was done using all variables in SDSS DR8 PhotoObjAll table. To identify variables that exhibit statistically significant difference between clockwise and counterclockwise galaxies, all variables were compared using an unpaired t-test. In the absence of an hypothesis regarding the direction of the expected difference, the two-tailed P values were used. 

When a large set of tests are performed, the probability that one of these tests provides statistically significant difference increases when the total number of tests gets higher. To avoid false positives, the Bonferroni correction \citep{goeman2014multiple} was applied to the t-test P values, so that the P value of each hypothesis is corrected for the total number of hypotheses analyzed in the experiment. Table~\ref{variables_manual} shows the variables that exhibit a Bonferroni-corrected statistical significance of $P<0.05$ for the difference between the values measured from clockwise galaxies and the values measured from counterclockwise galaxies.

%\begin{longtable}[ht]{lcccc}

\begin{table*}[ht]
\caption{Variables with Bonferroni-corrected statistically significant difference between clockwise and counterclockwise galaxies.}
\label{variables_manual}
\begin{center}
{
 \tiny 
\begin{tabular}{lcccc}
%\hline
\hline\hline
Variable       & Mean clockwise & Mean counterclockwise & t-test    & Bonferroni-corrected \\
                   &                        &                                   &    P       & t-test P   \\   
\hline
%\hline

psfMag\_g  & 19.73$\pm$0.003 & 19.75$\pm$0.003 & 0.00003 & 0.01 \\
psfMag\_z  & 18.39$\pm$0.003 & 18.4$\pm$0.003 & 0.00006 & 0.03 \\

fiberMag\_g  & 19.65$\pm$0.003 & 19.67$\pm$0.003 & 0.00003 & 0.01 \\
fiberMag\_r  & 18.89$\pm$0.003 & 18.91$\pm$0.003 & 0.00005 & 0.02 \\
fiberMag\_i & 18.48$\pm$0.003 & 18.5$\pm$0.003 & 0.00004 & 0.02 \\
fiberMag\_z  & 18.20$\pm$0.003 & 18.22$\pm$0.003 & 0.00004 & 0.02 \\

fiber2Mag\_g  & 20.39$\pm$0.003 & 20.4$\pm$0.003 & 0.00005 & 0.02 \\
fiber2Mag\_r  & 19.62$\pm$0.003 & 19.64$\pm$0.003 & 0.00007 & 0.03 \\
fiber2Mag\_i & 19.21$\pm$0.003 & 19.23$\pm$0.003 & 0.00007 & 0.03 \\
fiber2Mag\_z  & 18.91$\pm$0.003 & 18.94$\pm$0.003 & 0.00006 & 0.03 \\

petroMag\_g  & 17.53$\pm$0.004 & 17.55$\pm$0.003 & 0.00008 & 0.04 \\
petroMag\_i & 16.54$\pm$0.003 & 16.56$\pm$0.003 & 0.00005 & 0.02 \\

deVMag\_z  & 15.89$\pm$0.004 & 15.91$\pm$0.004 & 0.0001 & 0.04 \\

expMag\_z  & 16.39$\pm$0.004 & 16.41$\pm$0.004 & 0.00006 & 0.03 \\

psfFlux\_g  & 18.39$\pm$0.11 & 17.86$\pm$0.08 & 0.00006 & 0.03 \\
psfFlux\_r  & 34.47$\pm$0.17 & 33.5$\pm$0.15 & 0.00005 & 0.021 \\
psfFlux\_i  & 49.23$\pm$0.26 & 47.77$\pm$0.22 & 0.00002 & 0.007 \\
psfFlux\_z & 67.68$\pm$0.35 & 65.51$\pm$0.31 & $<10^{-5}$ & 0.001 \\

fiberFlux\_g  & 18.85$\pm$0.09 & 18.39$\pm$0.07 & 0.00005 & 0.02 \\
fiberFlux\_r  & 38.27$\pm$0.16 & 37.34$\pm$0.16 & 0.00004 & 0.02 \\
fiberFlux\_i  & 56.5$\pm$0.24 & 55.1$\pm$0.23 & 0.00003 & 0.01 \\

fiberFluxIvar\_g  & 34.82$\pm$0.06 & 35.26$\pm$0.06 & $<10^{-5}$ & 0.0001 \\
fiberFluxIvar\_r  & 14.16$\pm$0.02 & 14.38$\pm$0.03 & $<10^{-5}$ & $<10^{-5}$ \\
fiberFluxIvar\_i  & 6.42$\pm$0.01 & 6.5$\pm$0.01 & $<10^{-5}$ &  0.0002 \\

fiber2Flux\_g  & 9.67$\pm$0.05 & 9.42$\pm$0.04 & 0.00006 & 0.03 \\
fiber2Flux\_r  & 19.85$\pm$0.08 & 19.34$\pm$0.08 & 0.00003 & 0.01 \\
fiber2Flux\_i  & 29.46$\pm$0.13 & 28.69$\pm$0.12 & 0.00002 & 0.01 \\
fiber2Flux\_z  & 39.51$\pm$0.18 & 38.45$\pm$0.19 & 0.00007 & 0.03 \\

fiber2FluxIvar\_g  & 74.99$\pm$0.14 & 75.96$\pm$0.14 & $<10^{-5}$ & 0.0003 \\
fiber2FluxIvar\_r  & 30.6$\pm$0.056 & 31.1$\pm$0.06 & $<10^{-5}$ & $<10^{-5}$ \\
fiber2FluxIvar\_i & 14.02$\pm$0.03 & 14.19$\pm$0.03 & $<10^{-5}$ & 0.0006 \\

petroFluxIvar\_u  & 0.138$\pm$0.0003 & 0.14$\pm$0.0003 & 0.00003 & 0.01 \\
petroFluxIvar\_g  & 0.58$\pm$0.0014 & 0.59$\pm$0.001 & $<10^{-5}$ & $<10^{-5}$ \\
petroFluxIvar\_r  & 0.26$\pm$0.0007 & 0.27$\pm$0.0007 & $<10^{-5}$ & $<10^{-5}$ \\
petroFluxIvar\_i & 0.09$\pm$0.0002 & 0.092$\pm$0.0002 & $<10^{-5}$ & 0.0001 \\
petroFluxIvar\_z  & 0.0071$\pm$0.00002 & 0.0073$\pm$0.00002 & 0.00002 & 0.01 \\

deVAB\_g & 0.527$\pm$0.001 & 0.532$\pm$0.001 & $<10^{-5}$ & 0.0001 \\
deVAB\_r & 0.536$\pm$0.001 & 0.541$\pm$0.001 & $<10^{-5}$ & 0.0002 \\
deVAB\_i & 0.538$\pm$0.001 & 0.542$\pm$0.001 & $<10^{-5}$ & 0.001 \\

expAB\_g & 0.547$\pm$0.001 & 0.551$\pm$0.001 & $<10^{-5}$ & 0.0001 \\
expAB\_r & 0.553$\pm$0.001 & 0.558$\pm$0.001 & $<10^{-5}$ & 0.0002 \\
expAB\_i & 0.553$\pm$0.001 & 0.557$\pm$0.001 & $<10^{-5}$ & 0.001 \\

deVPhi\_u  & 91.15$\pm$0.2 & 88.64$\pm$0.2 & $<10^{-5}$ & $<10^{-5}$ \\
deVPhi\_g  & 91.13$\pm$0.2 & 87.58$\pm$0.2 & $<10^{-5}$ & $<10^{-5}$ \\
deVPhi\_r  & 91.55$\pm$0.2 & 88.15$\pm$0.2 & $<10^{-5}$ & $<10^{-5}$ \\
deVPhi\_i & 91.4$\pm$0.2 & 88.22$\pm$0.2 & $<10^{-5}$ & $<10^{-5}$ \\
deVPhi\_z  & 91.1$\pm$0.2 & 88.44$\pm$0.2 & $<10^{-5}$ & $<10^{-5}$ \\

expPhi\_u  & 91.06$\pm$0.2 & 88.53$\pm$0.2 & $<10^{-5}$ & $<10^{-5}$ \\
expPhi\_g  & 91.25$\pm$0.2 & 87.44$\pm$0.2 & $<10^{-5}$ & $<10^{-5}$ \\
expPhi\_r  & 91.49$\pm$0.2 & 87.87$\pm$0.2 & $<10^{-5}$ & $<10^{-5}$ \\
expPhi\_i & 91.38$\pm$0.2 & 87.94$\pm$0.2 & $<10^{-5}$ & $<10^{-5}$ \\
expPhi\_z  & 91.01$\pm$0.2 & 88.24$\pm$0.2 & $<10^{-5}$ & $<10^{-5}$ \\

u\_r  & -0.002$\pm$0.0005 & 0.001$\pm$0.0005 & 0.00002 & 0.01 \\

\hline
%\hline
\end{tabular}
}
\end{center}
\end{table*}

% http://skyserver.sdss.org/dr8/en/help/browser/browser.asp

The table shows that several different variables show a Bonferroni-corrected statistically significant difference between clockwise and counterclockwise galaxies. Some of these variables are the DeVaucouleurs fit position angle (deVPhi) variables computed from the u, g, r, i, and z bands, as well as the variables from the exponential fit position angle (expPhi) computed in the same bands. All of these variables show that the position angle measured for clockwise galaxies is, on average, smaller than the position angle measured for the galaxies that have counterclockwise patterns. The observation that the position angle is higher when measured on counterclockwise galaxies is in agreement with the analysis performed in \citep{sha16}, indicating that the source of the asymmetry observed in \citep{sha16} is not the human bias, but could be attributed to errors in the measurements performed by SDSS photometry pipeline. The SDSS `Stokes U' parameter measured in the r band (u\_r) also shows statistically significant difference. That observation also agrees with the differences in the `Stokes U' measured in the manually classified galaxies \citep{sha16}, although in this experiment statistically significant differences were only observed in the r band. The SDSS `Stokes U' parameter in SDSS is defined by $U=\frac{a-b}{a+b}\sin(2\phi)$, where {\it b} is the galaxy's minor axis, {\it a} is the major axis, and $\phi$ is the position angle \citep{abazajian2009seventh}. Since `Stokes U' is a function of the position angle, asymmetry in the measurement of the position angle could lead to differences in the measurement of the SDSS `Stokes U'.

%Other variables that separate between clockwise and counterclockwise galaxies are related to morphology, such as the PSF ellipticity profile (mE1PSF and mE2PSF) measured on bands u, g, r, i, and z, defined in SDSS as \citep{bernstein2002shapes}, and the size mRrCcPSF and mCr4PSF variables \citep{bernstein2002shapes} measured on these bands.

% http://www.sdss3.org/dr8/algorithms/
% http://egg.astro.cornell.edu/alfalfa/grads/hunt12/luke120626.html
% http://www.sdss3.org/dr8/algorithms/classify.php#photo_adaptive        clockwise more round.   counterclockwise more round
% http://skyserver.sdss.org/dr8/en/help/browser/description.asp?n=PhotoObjAll&t=U

The other photometric variables computed by SDSS pipeline that exhibit a statistically significant difference between clockwise and counterclockwise galaxies are related to the magnitude, showing that clockwise galaxies imaged by SDSS are brighter than counterclockwise galaxies. The PSF magnitude (psfMag) measurements on band g and z exhibit a statistically significant difference between clockwise and counterclockwise spiral galaxies, showing that clockwise galaxies observed through SDSS are brighter than counterclockwise galaxies. The same is observed by the comparison of the three arc seconds fiber magnitude (fiberMag) and two arcseconds fiber magnitude (fiber2Mag) measured in the g, r, i, and z bands measured by SDSS photometric pipeline. The de Vaucouleurs magnitude fit (deVMag) and exponential magnitude fit (expMag) measured on the z band, and the Petrosian magnitude measured on the g and i band also show that clockwise galaxies are brighter than counterclockwise galaxies. The PSF flux (psfFlux) and the three arc seconds fiber flux (fiberFlux) are higher for counterclockwise galaxies, with statistically significant difference observed in the g, r, i, and z bands. The inverse variance of the Petrosian flux (petroFluxIvar) shows that the measurement of counterclockwise galaxies is less noisy compared to clockwise galaxies. 

The model magnitude computed by SDSS pipeline also shows that in SDSS database, galaxies with clockwise handedness are brighter than galaxies with counterclockwise handedness, as shown in Table~\ref{model_mag}. Although the Bonferroni-corrected P values of these differences are not statistically significant, they are aligned with the magnitude differences shown in Table~\ref{variables_manual}.

\begin{table*}[ht]
\begin{center}
\caption{The mean, standard error of the mean, and t-test difference between model magnitude variables in clockwise and counterclockwise galaxies.}
\label{model_mag}
{\scriptsize 
\begin{tabular}{lcccc}
\hline\hline
Variable       & Mean clockwise & Mean counterclockwise & t-test  & Bonferroni-corrected \\
                   &                        &                                   &    P     & t-test P   \\   
\hline
u  & 18.93$\pm$0.004 & 18.95$\pm$0.004 & 0.01 & 1 \\
g  & 17.51$\pm$0.004 & 17.53$\pm$0.004 & 0.0003 & 0.16 \\
r  & 16.82$\pm$0.004 & 16.84$\pm$0.004 & 0.002 & 1 \\
i & 16.45$\pm$0.004 & 16.47$\pm$0.004 & 0.0005 & 0.25 \\
z  & 16.20$\pm$0.004 & 16.22$\pm$0.004 & 0.0004 & 0.2 \\

\hline
\end{tabular}
}
\end{center}
\end{table*}

The SDSS ratio between the major and minor axes measured using the exponential model fit (expAB) and the DeVaucouleurs model fit (DeVAB) measured in bands g, r, and i are significantly different between clockwise galaxies and counterclockwise galaxies in SDSS database, showing that according to the SDSS photometric pipeline counterclockwise galaxies tend to be more round than clockwise galaxies.

%That is evident by the values of variables such as psfFluxIvar, measured on the g, r, i, and z bands. psfFluxIvar measures the inverse variance of the PSF flux, and for bands g, r, i, and z the values measured from clockwise galaxies are higher than the values measured from counterclockwise galaxies. That is also true for inverse variance of the fiber flux magnitude (fiberFluxIvar), the inverse variance of the flux in two arcseconds fiber radius (fiber2FluxIvar), and inverse variance of the Petrosian flux (petroFluxIvar). However, in all cases the inverse variance measured on the u band is higher in counterclockwise galaxies compared to clockwise galaxies.

% The same is also observed for the inverse variance of the de Vaucouleurs magnitude model fit (deVFluxIvar) and the exponential magnitude model fit (expFluxIvar), but for these variables only the measurements in the u and z bands are statistically significant.

%FiberMag is the intensity in a three arcseconds diameter fiber radius. These variables measured on the u and g bands also show statistically significant difference between clockwise and counterclockwise galaxies, although these measurements are not related to the position angle. The difference between these variables shows that counterclockwise galaxies have a more luminous nucleus, and although the difference is small, it is statistically significant.

Given the high number of variables being tested, after the Bonferroni correction the P values of some of the differences might not be statistically significant. However, the differences between clockwise and counterclockwise galaxies are still in agreement with the statistically significant variables specified in Table~\ref{variables_manual}. For instance, Table~\ref{psfMag} shows the mean, standard error of the mean, and P values of the PSF magnitude on the i and r bands, as well as the fiberFlux, fiberFluxVar, and fiber2FluxIvar measured on the z band.

\begin{table*}[ht]
\begin{center}
\caption{Variables that their Bonferroni-corrected t-test is not statistically significant, but exhibit statistically significant t-test when measured in other bands.}
\label{psfMag}
{\scriptsize 
\begin{tabular}{lcccc}
\hline\hline
Variable       & Mean clockwise & Mean counterclockwise & t-test    & Bonferroni-corrected \\
                   &                        &                                   &    P       & t-test P   \\   
\hline
psfMag\_r  & 19.07$\pm$0.003 & 19.08$\pm$0.003 & 0.0004 & 0.21 \\
psfMag\_i  & 18.71$\pm$0.003 & 18.72$\pm$0.003 & 0.0001 & 0.07 \\
fiberFlux\_z  & 75.3$\pm$0.34 & 73.4$\pm$0.38 & 0.0002 & 0.1 \\
fiber2FluxIvar\_z  & 1.25$\pm$0.002 & 1.26$\pm$0.002 & 0.006 & 1 \\
fiberFluxIvar\_z  & 0.56$\pm$0.001 & 0.57$\pm$0.001 & 0.002 & 1 \\
\hline
\end{tabular}
}
\end{center}
\end{table*}

As the table shows, the Bonferroni-corrected P values of these variables do not show statistical significance, but the differences between the values measured from clockwise and counterclockwise galaxies are in agreement with the difference measured on the other bands. Therefore, while the difference in PSF magnitude of SDSS photometric pipeline measured in the r band does not exhibit strong Bonferroni-corrected statistical significance as shown by the PSF magnitude measured in the g band, the values measured in the r and i bands are clearly not in conflict with the PSF magnitude measured on the other bands.

\subsection{COMPARISON WITH PREVIOUS RESULTS}

The results shown in Table~\ref{variables_manual} were compared to previous experiments using galaxies that were annotated as spiral by Galaxy Zoo 2 \citep{sha16}. Among the variables in Table~\ref{variables_manual} showing asymmetry between clockwise and counterclockwise galaxies, just the `Stokes U' parameter measured in the r band also showed statistical significance in the Galaxy Zoo 2 annotated galaxies \citep{sha16}. That can be explained by the far smaller size of the set of Galaxy Zoo 2 spiral galaxies, not allowing the strong statistical significance of the difference that can be shown when the number of galaxies is high. Another reason can be the distribution of the Galaxy Zoo 2 galaxies used in the experiment. For instance, the Galaxy Zoo 2 galaxies that were used in that experiment were all in the RA range of (90$^o$,270$^o$), while the galaxies used in this study had no RA restriction. Also, the fact that the Galaxy Zoo galaxies were selected manually can introduce a bias driven by the human perception of the galaxies.

In addition to the Galaxy Zoo 2 galaxies, a previous experiment was also performed with a set of 10,281 automatically classified galaxies \citep{sha16}, showing that the handedness of the galaxy can be predicted by patterns of the photometric variables, but none of the single variables exhibited a Bonferroni-corrected statistical significance.  

While these datasets are much smaller and therefore less likely to identify variables with a statistically significant difference between clockwise and counterclockwise galaxies, if the asymmetry is consistent across these datasets the direction of the differences between the means should be in aligned. That is, if in one dataset the mean measured in clockwise galaxies is greater than the mean measured in the counterclockwise galaxies, it is expected that the mean will be also greater among clockwise galaxies in the other datasets. Therefore, if two datasets agrees, the sign of the difference between the means of clockwise and counterclockwise galaxies should be the same in both datasets. Table~\ref{comparison} shows the differences between the means of clockwise and counterclockwise galaxies of the different variables. The datasets are the large dataset described in Section~\ref{computer_dataset}, the Galaxy Zoo 2 galaxies described in \citep{sha16}, and the automatically classified galaxies described in \citep{sha16}. 
% Since the Galaxy Zoo 2 galaxies were analyzed using SDSS DR7, variables that are included in both data releases could be compared. 
The ``Stokes U''  parameter is statistically significant in both datasets and was therefore not compared.

\begin{table*}[ht]
\caption{The difference between the mean of different variables measured in clockwise galaxies and the mean measured in counterclockwise galaxies. The datasets are the dataset described in Section~\ref{computer_dataset}, the Galaxy Zoo 2 galaxies \citep{sha16}, and the dataset of automatically classified galaxies described in \citep{sha16}.}
\label{comparison}
\begin{center}
{
% \tiny
\scriptsize 
%\small
\begin{tabular}{lccc}
%\hline
\hline\hline
Variable       & Difference       & Difference            & Difference     \\
                   & (this dataset)  &  (Galaxy Zoo 2)     & (Automatic)   \\   
\hline
%\hline
deVPhi\_u  & 2.51 & 0.53 & 1.17 \\
deVPhi\_g  & 3.55 & 1.29 & 1.61 \\
deVPhi\_r  & 3.4  & 1.15 & 1.27 \\
deVPhi\_i &  3.19 & 1.07 & 0.53 \\
deVPhi\_z  & 2.64 & 1.35 & 0.19 \\

expPhi\_u  & 2.53 &  0.75 & 1.34 \\
expPhi\_g  &  3.81 & 1.08 & 2.56 \\
expPhi\_r  &  3.62 & 1.39 & 1.73 \\
expPhi\_i &  3.45 & 1.08 & 1.41 \\
expPhi\_z  & 2.77 & 0.95 & 0.91 \\

psfMag\_g  & -0.01 & 0.02 & -0.01 \\
psfMag\_z  &  -0.02 & 0.02 & -0.01 \\

fiberMag\_g  & -0.01 & 0.03 &  -0.01 \\
fiberMag\_r  & -0.01 &  0.03 & -0.01 \\
fiberMag\_i & -0.02 &  0.03 & -0.01 \\
fiberMag\_z  & -0.02  & 0.03 & -0.01 \\

fiber2Mag\_g  &  -0.01 & 0.03 & -0.01 \\
fiber2Mag\_r  &  -0.01 & 0.03 & -0.01 \\
fiber2Mag\_i &  -0.02 & 0.03 & -0.01 \\
fiber2Mag\_z  &  -0.02 & 0.03 & -0.005 \\

petroMag\_g  & -0.02 & 0.03  & -0.003 \\
petroMag\_i & -0.02  & 0.04   & -0.003 \\

deVMag\_z  & -0.02 & 0.05 & -0.003  \\

expMag\_z  & -0.02 & 0.05  & -0.004 \\

% psfFlux\_g  & 0.53 & & \\
% psfFlux\_r  &  0.92 & & \\
% psfFlux\_i  &  1.46 & & \\
% psfFlux\_z &  2.17 & & \\

deVAB\_g & -0.005 & -0.005  &  -0.002 \\
deVAB\_r & -0.005 & -0.004 &  -0.002 \\
deVAB\_i & -0.005 & -0.005 &  -0.003 \\

expAB\_g & -0.005 &-0.005  &  -0.002 \\
expAB\_r & -0.005 & -0.004 &  -0.001 \\
expAB\_i & -0.005 & -0.005 &  -0.001 \\

\hline
%\hline
\end{tabular}
}
\end{center}
\end{table*}

As the table shows, the differences between the means measured in clockwise and counterclockwise galaxies in the dataset described in Section~\ref{computer_dataset} is in full agreement with the dataset of automatically classified galaxies used in \citep{sha16}. In both datasets, if the mean measured on clockwise galaxies is greater than the mean measured on counterclockwise galaxies in the dataset described in Section~\ref{computer_dataset}, it is also be greater in the dataset of the automatically classified galaxies described in \citep{sha16}.   

The same agreement can also be observed between the dataset described in Section~\ref{computer_dataset} and the Galaxy Zoo 2 galaxies \citep{sha16} for the position angle variables (expPhi and devPhi) measured on the different bands, as well as the deVAB and expAB variables. A clear disagreement between the dataset described in Section~\ref{computer_dataset} and the Galaxy Zoo 2 galaxies is the variables related to magnitude. In the dataset described in Section~\ref{computer_dataset} the magnitude of counterclockwise galaxies is higher than the magnitude of clockwise galaxies, while in the Galaxy Zoo galaxies the magnitude of the clockwise galaxies is higher. The difference is consistent across all measurements related to the magnitude.  

The disagreement between the two datasets can be explained by the fact that the Galaxy Zoo 2 galaxies are all within the RA range of (90$^o$,270$^o$). As will be discussed in the next section and shown in Figure~\ref{mag_dif_ra}, in that specific RA range the clockwise galaxies have a higher magnitude than counterclockwise galaxies also in the dataset described in Section~\ref{computer_dataset}, showing a full agreement between all datasets.

\subsection{RESULTS USING MPA-JHU VARIABLES}
\label{mpajhu}

In addition to the photometric variables in the PhotoObjAll table of SDSS, another experiment compared the variables proposed by the MPA-JHU group \citep{brinchmann2004physical}, reflecting galaxy properties such as stellar mass, nebular oxygen abundance, and star formation rate. All 193 variables from the galSpecExtra table of SDSS DR8 were used in the same fashion the PhotoObjAll variables were analyzed. Since the variables are dependent on spectra, just 12668 clockwise galaxies and 12783 counterclockwise galaxies could be used. Table~\ref{mph-jhu} shows the mean, standard error, and t-test of the variables that exhibited the highest statistical significance of the difference between the values measured from clockwise galaxies and the values measured from counterclockwise galaxies.

\begin{table*}[ht]
\begin{center}
\caption{MPA-JHU variables measured from clockwise and counterclockwise galaxies.}
\label{mph-jhu}
{%\scriptsize
\small 
\begin{tabular}{lcccc}
\hline\hline
Variable       & Mean clockwise & Mean counterclockwise & t-test    & Bonferroni-corrected \\
                   &                        &                                   &    P       & t-test P   \\   
\hline
theta\_5  & -0.26$\pm$0.25 & -0.8$\pm$0.26 & 0.13 & 1 \\
theta\_6  & 0.72$\pm$0.41 & 1.58$\pm$0.44 & 0.15 & 1 \\
velDispChi2 & 1922.84$\pm$3.03 & 1930.39$\pm$3.26 & 0.09 & 1 \\
fracNSigma\_8 & 0.00041$\pm$0.00001 & 0.00044$\pm$0.00002 & 0.16 & 1 \\
fracNSigma\_9 & 0.0003$\pm$0.00001 & 0.0003$\pm$0.00001 & 0.14 & 1 \\
fracNSigma\_10 & 0.0002$\pm$0.00001 & 0.0002$\pm$0.00001 & 0.13 & 1 \\
fracNSigma\_8 & 0.00041$\pm$0.00001 & 0.00044$\pm$0.00001 & 0.16 & 1 \\
fracNSigma\_9 & 0.0003$\pm$0.00001 & 0.00033$\pm$0.00001 & 0.14 & 1 \\
fracNSigma\_10 & 0.00023$\pm$0.00001 & 0.00026$\pm$0.00001 & 0.13 & 1 \\
fracNSigLo\_7 & 0.00017$\pm$0.00001 & 0.00019$\pm$0.00001 & 0.09 & 1 \\
fracNSigLo\_8 & 0.00011$\pm$0.000004 & 0.00012$\pm$0.000004 & 0.09 & 1 \\
fracNSigLo\_9 & 0.00017$\pm$0.000004 & 0.00019$\pm$0.000004 & 0.059 & 1 \\
fracNSigLo\_10 & 0.00005$\pm$0.000003 & 0.000055$\pm$0.000004 & 0.057 & 1 \\
sn1\_i & 20.45$\pm$0.05 & 20.64$\pm$0.05 & 0.01 & 1 \\
sn2\_i & 21.34$\pm$0.05 & 21.51$\pm$0.06 & 0.035 & 1 \\
sn1\_r & 21.75$\pm$0.05 & 21.91$\pm$0.06 & 0.04 & 1 \\
sn2\_r & 23.42$\pm$0.06 & 23.57$\pm$0.06 & 0.09 & 1 \\
sn1\_g & 20.59$\pm$0.05 & 20.72$\pm$0.06 & 0.1 & 1 \\
sn2\_g & 22.72$\pm$0.06 & 22.84$\pm$0.06 & 0.18 & 1 \\

\hline
\end{tabular}
}
\end{center}
\end{table*}

Unlike the PhotoObjAll variables, none of the MPA-JHU variables showed a Bonferonni-corrected statistical significance. However, the number of galaxies that has MPA-JHU variables is much lower than the galaxies that have photometric information. The variables that showed the highest difference between clockwise and counterclockwise galaxies are the sn1 and sn2 variables, which are the $(\frac{S}{N})^2$ when g, r, and i are 20.2, 20.25, and 19.9, respectively, and measured in both spectrographs. Other variables that showed relatively higher difference are the fracNSigma\_X variables, which are the fraction of pixels more than X sigma relative to best-fit. The $\chi^2$ of the velocity dispersion fit (velDispChi2) and the coefficients for templates 5 and 6 fit (theta\_5 and theta\_6) were also among the variables that showed the highest difference, but were not statistically significant even without applying a Bonferroni correction.

\section{DISCUSSION}
\label{discussion}

The handedness is a noticeable morphological characteristic of a spiral galaxy. Since the handedness is dependent on the location of the observer, spiral galaxies with clockwise patterns are expected to be fully symmetric to spiral galaxies with counterclockwise handedness. The results described here clearly show that in SDSS database the handedness of a spiral galaxy can be predicted by the values of variables measured by SDSS photometric pipeline, showing a link between the handedness of the galaxy and its photometry.

Previous studies have shown that the human perception might not be a fully consistent tool to classify galaxies, especially when identifying the galaxy handedness \citep{land2008galaxy}. The results of this experiment are shown with galaxies classified in a completely automatic process, and without human intervention, guaranteeing that the results are not driven by a bias in the human perception. 

The results show a statistically significant higher number of galaxies with clockwise handedness compared to galaxies with counterclockwise handedness in the SDSS database. These results are in agreement with previous studies using manually classified galaxies \citep{sha16,shamir2012handedness,hoehn2014characteristics}, but other studies using a smaller number of galaxies \citep{sha16} or manually classified galaxies \citep{land2008galaxy} do not show statistically significant preference for a certain handedness. The differences between some of the measurements are statistically significant, and were deduced using a fully automatic process, leading to the conclusion that the SDSS photometric pipeline is sensitive to the handedness of the galaxy, and provided different photometric measurements for galaxies of different handedness.

Reasons for the asymmetry of the position angle can be the way the position angle is measured. Unlike elliptical galaxies, the morphology of a spiral galaxy can be asymmetric along the major axis, and that asymmetry can be also dependent on the handedness, leading to an error in the measurement of the position angle.

\begin{figure}[ht]
%\figurenum{<text>}
\includegraphics[scale=0.7]{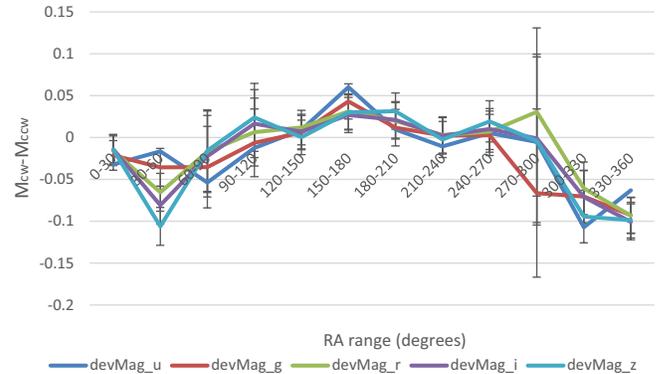}
%\plottwo{<epsfile>}{<epsfile>}
\caption{Differences between the de Vaucouleurs magnitude of clockwise galaxies and the de Vaucouleurs magnitude of counterclockwise galaxies in different RA ranges in the u, g, r, i, and z bands.}
\label{mag_dif_ra}
\end{figure}

The mean magnitude of clockwise SDSS galaxies in a certain part of the sky is expected to be approximately the same as the mean magnitude of counterclockwise SDSS galaxies in the same sky region. Therefore, even if the measurements of the magnitude change in different sky regions due to atmospheric or other effects, the measurements should be on average the same for clockwise and counterclockwise galaxies, as all galaxies are in the same sky region. Figure~\ref{mag_dif_ra} shows the difference between the mean de Vaucouleurs magnitude of galaxies with clockwise handedness and the mean de Vaucouleurs magnitude of galaxies with counterclockwise handedness in different RA sectors of 30$^o$. The graph shows that the difference is smaller around the RA range of (30$^o$,60$^o$), increases until peaking in the RA range of (180$^o$,210$^o$), and then starting to decrease. The graph shows a symmetric pattern centered at around the RA range of (180$^o$,210$^o$).

The inverse Pearson correlation between the cosine of the RA and the measured difference between the g magnitude of clockwise galaxies and g magnitude of counterclockwise galaxies is $\sim$0.868. Given that the sample size is 12, the two-tailed probability to have such correlation by chance is $(P<0.00025)$. Clearly, the other bands are strongly correlated with the g band, and therefore it is expected that all bands exhibit a strong correlation with the cosine of the RA. The Pearson correlation of the u, r, i, and z bands with the cosine of the RA is 0.75, 0.731, 0.787, and 0.761, respectively. The two-tailed probability for having such correlation by chance is 0.005, 0.007, 0.002, and 0.004, respectively.

\begin{figure}[ht]
%\figurenum{<text>}
%\epsscale{1}
\includegraphics[scale=0.7]{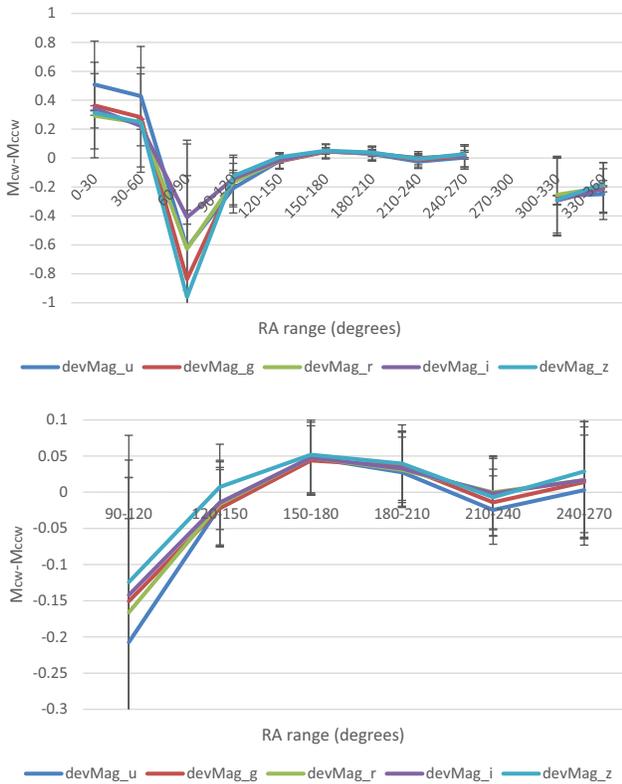}
%\plottwo{<epsfile>}{<epsfile>}
\caption{Differences between the de Vaucouleurs magnitude of clockwise galaxies and the de Vaucouleurs magnitude of counterclockwise galaxies in different RA ranges in the u, g, r, i, and z bands. The galaxies are taken from the dataset of automatically classified galaxies used in \citep{sha16}. The bottom graph shows just the RA range of (90$^o$,270$^o$), where the population of the galaxies is higher and therefore the standard error is lower.}
\label{mag_dif_ra_small}
\end{figure}

These results were compared to the differences in the magnitude of clockwise and counterclockwise galaxies in the two smaller datasets used in \citep{sha16}. Figure~\ref{mag_dif_ra_small} shows the difference between the de Vaucouleurs magnitude of clockwise galaxies and the de Vaucouleurs magnitude of counterclockwise galaxies in different RA ranges using the dataset of automatically annotated galaxies used in \citep{sha16}. Outside the RA range of (90$^o$,270$^o$) the population of galaxies is low, making the asymmetry large in some of these weakly populated RA ranges, and therefore the figure shows the differences in the RA range of (90$^o$,270$^o$). 

As the figure shows, within the RA range of (90$^o$,270$^o$), the dataset used in \citep{sha16} and the dataset used in this paper are largely in agreement. A noticeable exception is the i and z bands in the (90$^o$,120$^o$) RA range, but that range also has a larger standard error. 

\begin{figure}[h!]
%\figurenum{<text>}
%\epsscale{1}
\includegraphics[scale=0.7]{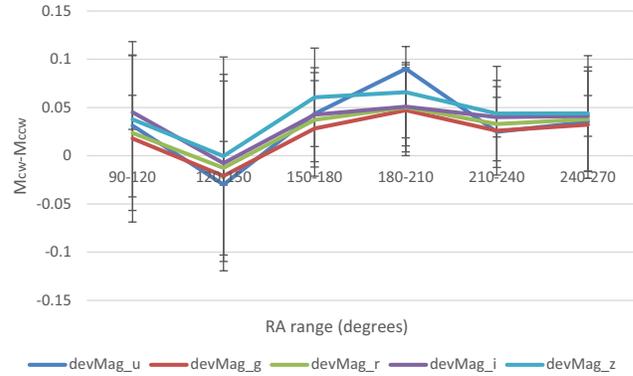}
%\plottwo{<epsfile>}{<epsfile>}
\caption{Differences between the de Vaucouleurs magnitude of clockwise galaxies and the de Vaucouleurs magnitude of counterclockwise galaxies in different RA ranges in the u, g, r, i, and z bands. The galaxies are taken from the dataset of Galaxzy Zoo 2 galaxies used in \citep{sha16}. The graph shows the RA range of (90$^o$,270$^o$), as the galaxies of that dataset are within that RA range.}
\label{mag_dif_ra_gz2}
\end{figure}

Figure~\ref{mag_dif_ra_gz2} shows the same using the dataset of Galaxy Zoo 2 galaxies used in \citep{sha16}. The galaxies of that dataset are within the RA range of (90$^o$,270$^o$). The differences between the clockwise and counterclockwise galaxies in the Galaxy Zoo 2 dataset are also largely aligned with the results of the dataset used in this study. 

The handedness of a spiral galaxy is a crude binary descriptor, and there is no known atmospheric or other effect that can make a galaxy with clockwise handedness seem to have counterclockwise patterns. Also, the ratio between the frequency of clockwise and counterclockwise galaxies is different in different parts of the sky covered by SDSS, and therefore if such asymmetry exists, it is sensitive to the region of the sky. The vast majority of the galaxies used in this study do not have spectra, making it difficult to separate gravitationally interacting superstructures such as the Sloan Great Wall \citep{gott2005map}. 

Measurements of the differences between the magnitude of clockwise and counterclockwise galaxies are taken using galaxies imaged in the same parts of the sky, but are separated by their handedness. The fact that the galaxies are imaged at the same sky region guarantees that the differences are not driven by different atmospheric effects. 

Reasons for such asymmetry in the SDSS database could be related to a systematic measurement bias in the photometry pipeline. Related observations have also been related to large-scale asymmetry in the local universe \citep{longo2011detection,shamir2012handedness,gullu2013observable}. While that explanation conflicts with some current cosmological assumptions, current cosmological models might not be complete \citep{kroupa2012dark,kragh2013most}, and the cosmological principle is challenged by the observation of possible large structures \citep{clowes2013structure,horvath2013largest}, exceeding the size limit of structures that do not violate the homogeneity aspect of the cosmological principle \citep{yadav2010fractal}. The observed asymmetry can also be the result of local galaxy properties such as internal extinction along the galaxy arms, which can be linked to the radial velocity. The results described in this study propose questions for further investigation into the source and cause of the observed asymmetry.

The primary downside of the results described in this paper is that they rely on a single source, which is the Sloan Digital Sky Survey. While it is difficult to identify reasons that can cause such asymmetry, photometric pipelines and imaging systems are complex, and can be vulnerable to errors of kinds that have not been identified in the past. Further studies will use equivalent databases such as PanStarrs, as well as far larger databases such as the future LSST to test whether the observation is consistent across databases, or specific to the widely used SDSS database.

%The results of this study also propose a new approach the studying the large-scale structure of the universe. While redshift surveys have been used to study the large-scale structure of the universe by analyzing galaxy population \citep{colless2000redshift}, in this study the asymmetry between clockwise and counterclockwise galaxies is considered as a possible probe to study the properties of the universe in its large scale. 

%There is no known effect that can make a galaxy with clockwise handedness seem to have counterclockwise patterns, but the observation that the position angle has a systematic bias that is sensitive to the handedness of the galaxy demonstrates that some of SDSS measurements are affected by the handedness. That leaves the possibility that the observed asymmetry is not necessarily of a cosmological nature, but can also be the product of bias in SDSS pipeline.

\section{Acknowledgments}

I would like to thank the anonymous reviewer for the comments that helped to improve it. The research was supported in part by NSF grant IIS-1546079.

\bibliographystyle{pasa-mnras}
\bibliography{galaxy_rotation_assym2}

\vspace{2cm}

\appendix{Appendix: Sample dataset of 800 galaxies}
\label{sample_galaxies}

The following Figures~\ref{cw16x25_1} and~\ref{ccw16x25_1} show the 400 sample clockwise galaxies and 400 sample counterclockwise galaxies, respectively.

\begin{figure*}
%\figurenum{<text>}
%\epsscale{1}
\includegraphics[scale=0.3]{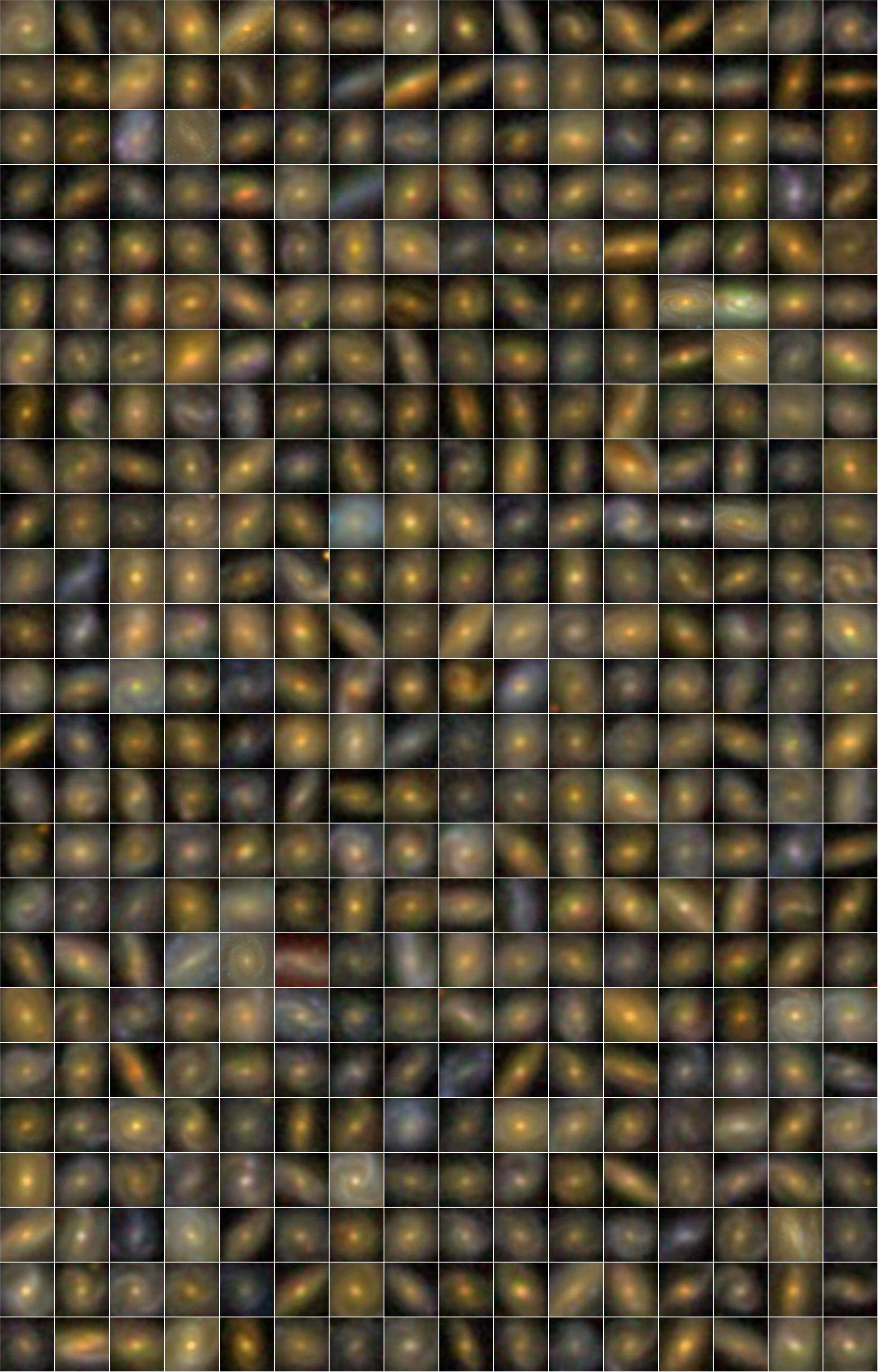}
%\plottwo{<epsfile>}{<epsfile>}
\caption{The sample dataset of four hundred galaxies with clockwise handedness.}
\label{cw16x25_1}
\end{figure*}

\begin{figure*}
%\figurenum{<text>}
%\epsscale{1}
\includegraphics[scale=0.3]{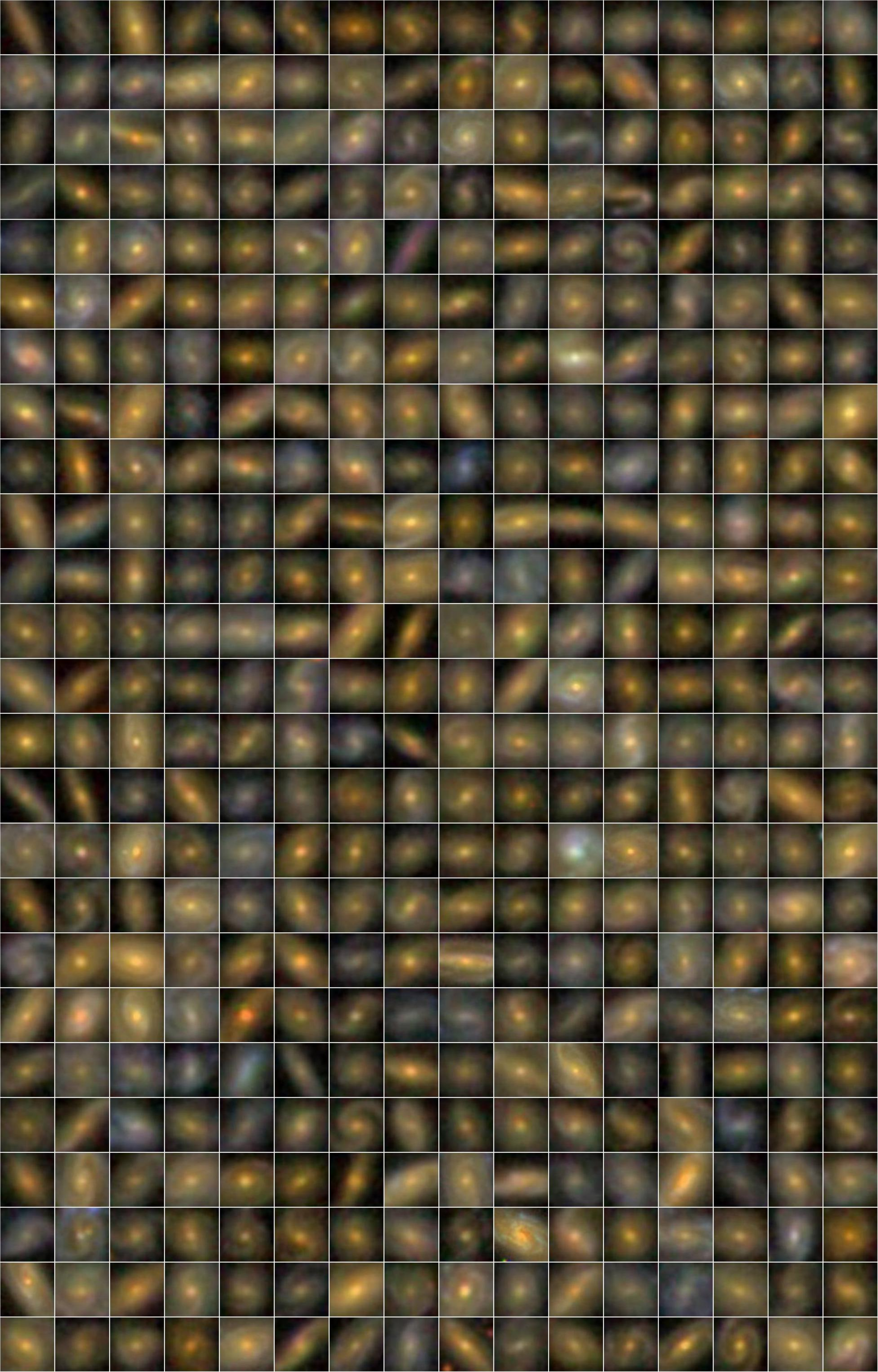}
%\plottwo{<epsfile>}{<epsfile>}
\caption{The sample dataset of four hundred galaxies with counterclockwise handedness.}
\label{ccw16x25_1}
\end{figure*}

%% the file, uncomment the \plotone, \plottwo, and
%% If you need a layout that cannot be achieved with \plotone or
%% \plottwo, you can invoke the graphicx package directly with the
%% \includegraphics command or use \plotfiddle. For more information,
%% please see the tutorial on "Using Electronic Art with AASTeX" in the
%% documentation section at the AASTeX Web site, http://aastex.aas.org/
%%
%% The examples below also include sample markup for submission of
%% supplemental electronic materials. As always, be sure to check
%% the instructions to authors for the journal you are submitting to
%% for specific submissions guidelines as they vary from

%% This example uses \plotone to include an EPS file scaled to
%% 80% of its natural size with \epsscale. Its caption
%% has been written to indicate that additional figure parts will be
%% available in the electronic journal.

\clearpage

%% If you use the table environment, please indicate horizontal rules using
%% Do not put multiple tabular environments within a single table.
%% The optional \label should appear inside the \caption command.

\end{document}